\documentclass[prl,twocolumn,preprintnumbers,amsmath,amssymb,floats,citeautoscript,nobalancelastpage,superscriptaddress]{revtex4}

\usepackage{graphicx}
\usepackage{dcolumn}
\usepackage{bm}
\usepackage{times}
\usepackage{txfonts}
\usepackage{color}
\newcommand{\average}[1]{\ensuremath{\langle#1\rangle}}

\begin{document}

\title{Anticorrelation between polar lattice instability and superconductivity in the Weyl semimetal candidate MoTe$_{2}$}

\author{H.~Takahashi}
\affiliation{Department of Applied Physics and Quantum Phase Electronics Center (QPEC), University of Tokyo, Tokyo 113-8656, Japan}
\author{T.~Akiba}
\affiliation{Department of Applied Physics and Quantum Phase Electronics Center (QPEC), University of Tokyo, Tokyo 113-8656, Japan}
\author{K.~Imura}
\affiliation{Department of Physics, Nagoya University, Nagoya 464-8602, Japan}
\author{T.~Shiino}
\affiliation{Department of Physics, Nagoya University, Nagoya 464-8602, Japan}
\author{K.~Deguchi}
\affiliation{Department of Physics, Nagoya University, Nagoya 464-8602, Japan}
\author{N.~K.~Sato}
\affiliation{Department of Physics, Nagoya University, Nagoya 464-8602, Japan}
\author{H.~Sakai}
\affiliation{Department of Applied Physics and Quantum Phase Electronics Center (QPEC), University of Tokyo, Tokyo 113-8656, Japan}
\affiliation{Department of Physics, Osaka University, Toyonaka, Osaka 560-0043, Japan}
\affiliation{PRESTO, Japan Science and Technology Agency, Kawaguchi, Saitama 332-0012, Japan}
\author{M.~S.~Bahramy}
\affiliation{Department of Applied Physics and Quantum Phase Electronics Center (QPEC), University of Tokyo, Tokyo 113-8656, Japan}
\affiliation{RIKEN Center for Emergent Matter Science (CEMS), Wako 351-0198, Japan.}
\author{S.~Ishiwata}
\affiliation{Department of Applied Physics and Quantum Phase Electronics Center (QPEC), University of Tokyo, Tokyo 113-8656, Japan}
\affiliation{PRESTO, Japan Science and Technology Agency, Kawaguchi, Saitama 332-0012, Japan}


\begin{abstract}
The relation between the polar structural instability and superconductivity in a Weyl semimetal candidate MoTe$_{2}$ has been clarified by finely controlled physical and chemical pressure. The physical pressure as well as the chemical pressure, i.e., the Se substitution for Te, enhances the superconducting transition temperature $T_{\rm c}$ at around the critical pressure where the polar structure transition disappears. From the heat capacity and thermopower measurements, we ascribe the significant enhancement of $T_{\rm c}$ at the critical pressure to a subtle modification of the phonon dispersion or the semimetallic band structure upon the polar-to-nonpolar transition. On the other hand, the physical pressure, which strongly reduces the interlayer distance, is more effective on the suppression of the polar structural transition and the enhancement of $T_{\rm c}$ as compared with the chemical pressure, which emphasizes the importance of the interlayer coupling on the structural and superconducting instability in MoTe$_{2}$. 
\end{abstract}


\maketitle
Inversion symmetry breaking and polar structural instability in metallic compounds have received growing attention as a key factor for exploring exotic electronic states \cite{1,2,3,4,5,7,33,22,51}. For superconductivity, breaking the inversion symmetry yields unique features by lifting the spin degeneracy through the spin-orbit coupling, for instance, the enhanced upper critical field due to spin-valley locking and the mixing of the singlet and triplet pair states with the Majorana fermions at edge channels \cite{3,4,5,7}. Despite the considerable interest, the detailed effect of the inversion symmetry breaking on the superconductivity remains elusive because of the lack of the superconducting compounds, of which inversion symmetry breaking can be tuned by external parameters such as pressure.

Given the semimetallic band dispersion along with the inversion symmetry breaking and spin-orbit coupling, there may exist topologically protected crossing points at the vicinity of the Fermi level, which allows the emergence of the Weyl fermion as a low-energy excitation \cite{43,44,45}. Because of the fundamental importance and the technological potential, the search for a new Weyl semimetal (WSM) has currently become of great concern. Recently, transition-metal dichalcogenides T$_{\rm d}$-(Mo,W)Te$_{2}$ with inversion symmetry breaking have attracted much attention because of their potential to be a type-I\hspace{-.1em}I WSM, which is characterized by a pair of Weyl points connected by the gapless surface states called Fermi arcs \cite{35,10,32,36}. Furthermore, T$_{\rm d}$-(Mo,W)Te$_{2}$ show distinct electronic properties such as extremely large magnetoresistance and pressure-enhanced superconductivity, implying subtle sensitivity of the electronic state to external fields or pressure \cite{42,30,31}. To be noted here is that MoTe$_{2}$ shows a transition from a high-temperature nonpolar ($P2_{1}/m$) 1T' structure to the low temperature polar ($Pnm2_{1}$) T$_{\rm d}$ structure at $T_{\rm s}$ $\sim$250 K [see Fig. 1(c)] \cite{15}, which can be suppressed by external pressure or chemical substitution \cite{16, 22}. It has been reported that the superconducting transition temperature $T_{\rm c}$ of about 0.1 K in the polar phase increases to go beyond 5 K as the polar transition is suppressed by external pressure  \cite{16}. A similar enhancement of $T_{\rm c}$ upon the suppression of the polar structural distortion in MoTe$_{2}$ has been found by the S substitution for Te \cite{17}. Considering the fact that exotic superconductivity expected in the WSM state emerges only in the T$_{\rm d}$ (polar) phase \cite{46,47,48}, it is of vital importance to clarify the detailed relation between the polar structural transition and the superconductivity in MoTe$_{2}$, which remains elusive because of the difficulty in the physical property measurements under pressure. 

In this Rapid Communication, we adopt not only the physical pressure but the chemical pressure, that is the Se substitution for Te, to establish the phase diagram of MoTe$_{2}$ as a function of temperature and pressure. Upon the application of physical and chemical pressures, $T_{\rm c}$ of 0.1 K in MoTe$_{2}$ increases significantly by a factor of more than 20 with replacing the T$_{\rm d}$ (polar) phase by the 1T' (nonpolar) phase. To clarify the role of the polar lattice distortion in the superconductivity, we measured the heat capacity and thermopower of the polar and nonpolar phases in Mo(Te$_{1-x}$Se$_{x}$)$_{2}$ and found that the superconductivity is sensitive to the subtle modification of the phonon dispersion or the Fermi surface topology. In addition, the importance of the interlayer coupling both on the structural and superconducting transitions in MoTe$_{2}$ is discussed from the viewpoint of the anisotropic lattice change under pressure.

Single crystals of MoTe$_{2}$ were prepared by the chemical vapor transport method as reported in Ref. \:\onlinecite{18}. Polycrystalline samples of Mo(Te$_{1-x}$Se$_{x}$)$_{2}$ with $x=0, 0.05, 0.1, 0.15, 0.2$ were synthesized by solid state reaction in evacuated quartz tubes. Further details of the experimental methods are provided in the Supplemental Material \cite{40}.   

\begin{figure}[t]
\begin{center}
\includegraphics[width=7cm]{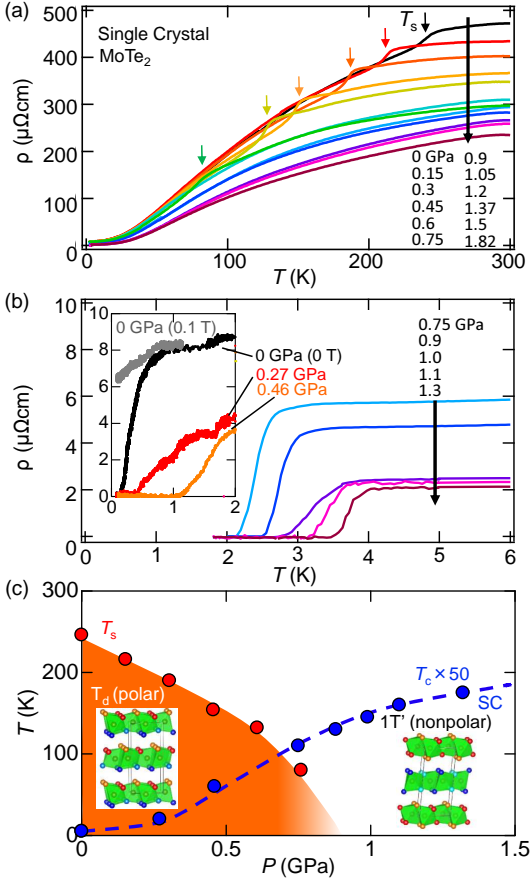}
\caption{(color online). (a),(b) The electrical resistivity for the single crystalline MoTe$_{2}$ at various pressures. Inset of Fig. 1(b) shows the electrical resistivity at pressures up to 0.46 GPa. The superconductivity at 0 GPa is suppressed by applying the magnetic field of 0.1 T as plotted by the gray line. (c) Phase diagram of the crystal structure and superconductivity. Red and blue circles represent the structural transition temperature $T_{\rm s}$ and superconducting transition temperature $T_{\rm c}$, respectively. Inset shows the crystal structure of the T$_{\rm d}$ (polar) phase and the 1T' (nonpolar) phase.}
\end{center}
\end{figure}

Figures 1(a) and 1(b) show the temperature dependence of $\rho$ for the single-crystalline MoTe$_{2}$ at various pressures. The residual resistivity ratio (RRR$\sim $ 60) is larger than that of Ref. \:\onlinecite{16} (RRR$\sim$ 36), ensuring the quality of our single crystal. At ambient pressure, the $\rho$-$T$ curve shows the anomaly reflecting the structural transition at $T_{\rm s}$ $\sim$250 K as well as the zero resistivity indicating the superconducting transition at $T_{\rm c}$ $\sim$0.1 K. With increasing pressure, $T_{\rm s}$ decreases so that it becomes zero at the critical pressure $P_{\rm c}$ of 0.75 GPa, whereas $T_{\rm c}$ increases systematically with increasing pressure up to 1.3 GPa. The point here is that the slope of $T_{\rm c}$ as a function of $P$ is maximized ($dT_{\rm c}/dP\sim$ 4 K/GPa) at the vicinity of the critical pressure $\leq P_{\rm c}$ (0.27 GPa $\leq P \leq$ 0.75 GPa) separating the T$_{\rm d}$ phase and the 1T' phase at the lowest temperature. Note that $T_{\rm c}$ changes significantly but continuously around $P_{\rm c}$. The smeared increase in $T_{\rm c}$ with increasing $P$ may reflect the fact that the structural transition is of the first order, giving rise to the phase coexistence at low temperatures at the vicinity of $P_{\rm c}$. \cite{22}

\begin{figure}[t]
\begin{center}
\includegraphics[width=7cm]{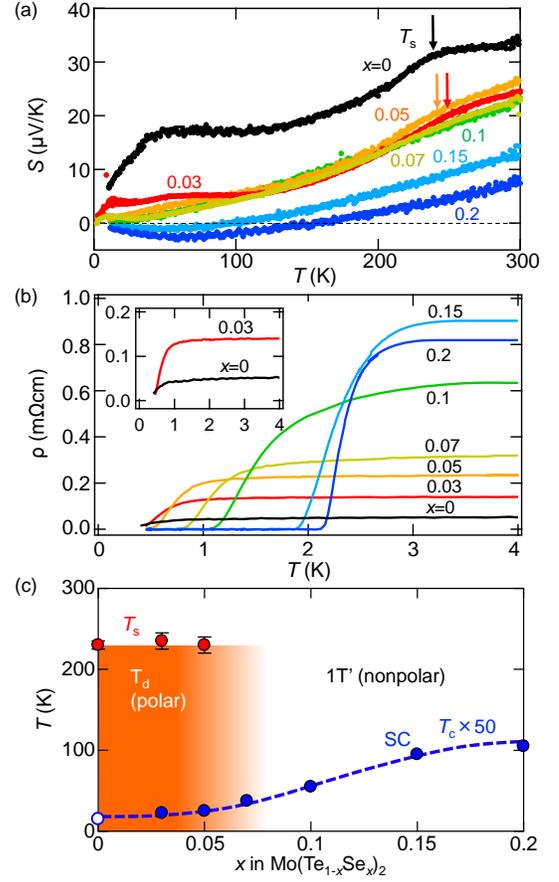}
\caption{(color online). (a) The Seebeck coefficient and (b) resistivity for the polycrystalline samples of Mo(Te$_{1-x}$Se$_{x}$)$_{2}$ below 4 K. (c) Phase diagram of the structure and superconductivity. Structural transition temperature $T_{\rm s}$ (red circles) and superconducting transition temperature $T_{\rm c}$ (blue circles) are determined by the Seebeck coefficient and resistivity, respectively.}
\end{center}
\end{figure}

The relation between the polar structural instability and superconductivity is also studied by the Se substitution for Te, which can be regarded as a chemical pressure effect as will be discussed later. Mo(Te$_{1-x}$Se$_{x}$)$_{2}$ with $0\leq x\leq0.2$ crystallizes in the monoclinic 1T' structure at room temperature. As shown in Fig. 2(a), a weak anomaly corresponding to the structural transition at $T_{\rm s}$ is discernible in $S$ $(T)$ for Mo(Te$_{1-x}$Se$_{x}$)$_{2}$ with $0\leq x\leq 0.05$, which is smeared in $\rho$ $(T)$ but discernible by x-ray diffraction at low temperatures (see Fig. S1 in the Supplemental Material) \cite{40}. Regarding the $x$ dependence of $S$, the magnitude of $S$ at room temperature tends to decrease with increasing $x$, while $T_{\rm s}$ remains constant. The temperature dependence of $\rho$ below 4 K for the polycrystalline Mo(Te$_{1-x}$Se$_{x}$)$_{2}$ is shown in Fig. 2(b). The compound with $x=0$ shows the decrease of $\rho$ around 0.3 K, which is reminiscent of superconductivity. For the compounds with $0.03\leq x\leq 0.2$, the superconductivity emerges at $T_{\rm c}$ above 0.3 K. From the result of $\rho$ $(T)$ and $S$ $(T)$, the structural and superconducting phase diagram of Mo(Te$_{1-x}$Se$_{x}$)$_{2}$ is obtained as shown in Fig. 2(c). Interestingly, $T_{\rm c}$ increases significantly as the T$_{\rm d}$ phase is replaced by the 1T' phase, just like the case in MoTe$_{2}$ under physical pressure.  

\begin{figure}[t]
\begin{center}
\includegraphics[width=7cm]{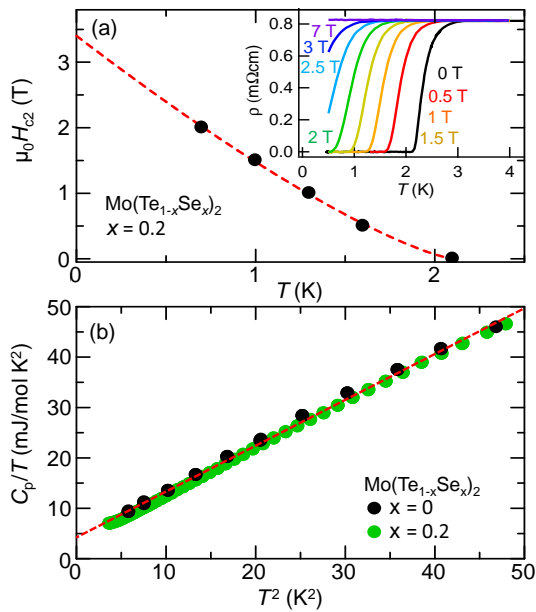}
\caption{(color online). (a) Temperature dependence of upper critical field $H_{\rm c2}$ for $x=0.2$. The red curve is the best fit of the equation $H_{\rm c2}=H^{*}_{\rm c2}(1-T/T_{\rm c})^{1+\alpha }$ to the experimental data. The inset shows the temperature dependence of the resistivity under magnetic fields up to 7 T. (b) $T^{2}$ dependence of $C_{\rm p}/T$ of polycrystalline samples of $x=0$ (black circles) and $x=0.2$ (green circles). The red dotted line shows the fitting of the experimental data with the equation $C_{\rm p}/T=\gamma +\beta T^{2}$.}
\end{center}
\end{figure}

To get insight into the superconducting properties in the 1T' phase, we measured $\rho$ for Mo(Te$_{1-x}$Se$_{x}$)$_{2}$ with $x=0.2$ under various external magnetic fields $H$ as shown in the inset of Fig. 3(a). The change in the zero-resistance-point $T_{\rm c}$ with increasing $H$ is summarized as the phase diagram of the upper critical field $H_{\rm c2}(T)$ in Fig. 3(a). The curve of $H_{\rm c2}(T)$ becomes concave upward close to $T_{\rm c}$ ($H=0$), which deviates from the one expected from the Werthamer-Helfand-Hohenberg theory based on the single-band model. A similar behavior has been reported for 1T'-(Mo,W)Te$_{2}$ under high pressure and 2H-NbSe$_{2}$ \cite{16,29,30}.

Next, we compare the normal state properties of polar and nonpolar phases in Mo(Te$_{1-x}$Se$_{x}$)$_{2}$ with $x=0$ and $x=0.2$, respectively. To analyze the phonon properties and electronic density of states, we measured the temperature dependence of the specific heat $C_{\rm p}$. Figure 3(b) shows $C_{\rm p}/T$ as a function of $T^{2}$ above 2 K. The data of the polar ($x=0$) and nonpolar ($x=0.2$) samples are almost identical with each other, suggesting that both samples have similar Sommerfeld constant $\gamma $ and Debye temperature $\Theta _{\rm D}$. $\gamma $ and $\Theta _{\rm D}$ are evaluated using the equation $C_{\rm p}/T=\gamma +\beta T^{2}$ ($\beta$ is the lattice contribution to the specific heat), which yields $\gamma = 3.7$ mJ/mol K$^{2}$ and $\beta=0.923$ mJ/mol K$^{4}$. $\Theta _{\rm D}$ can be calculated with $\beta=(12/5)\pi ^{4}NR\Theta _{\rm D}^{3}$ ($R=8.314$ J/mol K and $N=3$) to be $180$ K.

Here we discuss the possible origin of the large enhancement of $T_{\rm c}$ upon the structural transition from the polar to the nonpolar phase in Mo(Te$_{1-x}$Se$_{x}$)$_{2}$. Using the evaluated $\Theta _{\rm D}$, the electron-phonon coupling strength $\lambda _{\rm p}$ can be calculated with the equation of the McMillan formula \cite{24,25},
\begin{equation}
\begin{aligned}
\lambda _{\rm p}=\frac{\mu^{*} {\rm ln}(1.45T_{\rm c}/\Theta _{\rm D})-1.04}{1.04+{\rm ln}(1.45T_{\rm c}/\Theta _{\rm D})(1-0.62\mu^{*})}.
\end{aligned}
\label{eq:ep2}
\end{equation} 
By assuming the Coulomb pseudopotential $\mu ^{*}=0.1$ (commonly accepted value), $\lambda _{\rm p}$ is evaluated to be 0.31 and 0.52 for $x=0$ and $x=0.2$, respectively.
These values are similar to that in Ref. \:\onlinecite{17}, indicating that both compounds are weak-coupling superconductors.
In general, $\lambda _{\rm p}$ can be qualitatively expressed as
\begin{equation}
\begin{aligned}
\lambda _{\rm p}=\Sigma _{\rm i}\frac{\average{I^{2}_{\rm i}}D_{\rm i}(E_{F})}{M_{\rm i}\average{\omega ^{2}_{\rm i}}},
\end{aligned}
\label{eq:ep2}
\end{equation}  
where $\average{I^{2}_{\rm i}}$, $D_{\rm i}(E_{\rm F})$, $M_{\rm i}$, and $\average{\omega ^{2}_{\rm i}}$ are the mean-squared electron-phonon coupling matrix element averaged over the Fermi surface, electronic density of states at the Fermi surface, atomic mass, and averaged squared phonon frequency of $i$th atoms in the unit cell, respectively \cite{26,27,28}. Since $\gamma$ and $\Theta _{\rm D}$ estimated from the specific heat measurements remain almost intact upon the Se doping, $D_{\rm i} (E_{\rm F})$ and $\omega_{\rm i}$ seem to be less dependent on the crystal structure and Se content $x$. In fact, the structural transition between the 1T' and the T$_{\rm d}$ phases has little impact on $D_{\rm i} (E_{\rm F})$ as confirmed by the first-principles calculations (see Fig. S2 in the Supplemental Material) \cite{40}. As for the contribution to $\lambda _{\rm p}$ from the atomic mass, the reduction of $M_{\rm i}$ from 351.14 g/mol ($x=0$) to 331.7 g/mol ($x=0.2$) by the Se substitution is too small to be considered. Given the positive correlation between $\lambda _{\rm p}$ and $T_{\rm c}$, it is presumable that the enhancement of $T_{\rm c}$ upon the transition from the T$_{\rm d}$ to the 1T' phase is associated with the increase in $\average{I^{2}_{\rm i}}$. The change in $\average{I^{2}_{\rm i}}$ accompanied by the structural change may be attributed to the modification of band topology. As shown in Fig. 2(a), the possible change in the band topology is supported by the sign change in $S$ at low temperatures for Mo(Te$_{1-x}$Se$_{x}$)$_{2}$ at $x=0.1-0.15$. Given that the Se ion and the Te ion are isovalent, the change in $S$ by the Se substitution is attributable not to the shift in the chemical potential but to a subtle change in the semimetallic band structure near the Fermi level. Thus the change in the topology of the semimetallic band structure is worth considering as a factor for the reduction of $\average{I^{2}_{\rm i}}$.  To see whether or not the change in the band topology is associated with the spin splitting, further experiments such as spin-polarized photoemission spectroscopy are necessary. 

On the other hand, $\lambda_{\rm p}$ can be calculated using the Eliashberg electron-phonon spectral function $\alpha^{2}F(\omega )$, expressed as $\lambda _{\rm p}=\int_{0}^{\infty }$d$\omega \alpha^{2}F(\omega )/\omega $. Recently, it has been reported that a transverse acoustic phonon mode corresponding mainly to the Te-Te interlayer vibrations in the 1T' phase of WTe$_{2}$ has a significant contribution to $\lambda_{\rm p}$ through the large density of $\alpha^{2}F(\omega )$\cite{50}. This result suggests that the small change of the interlayer distance and/or the softening of the interlayer Te-Te vibration modes upon the polar to nonpolar structure transition substantially affect $T_{\rm c}$ in MoTe$_{2}$ as well, even if the overall phonon modes remain almost the same \cite{41,49}. The significant correlation between the interlayer distance and the superconductivity is discussed below.
      
While the qualitative features of the superconductivity under physical and chemical pressure are similar in the sense that $T_{\rm c}$ in the 1T' phase is much higher than that in the T$_{\rm d}$ phase, there are quantitative differences with regard to the changes in $T_{\rm s}$ and $T_{\rm c}$ as a function of each pressure as shown in Figs. 1(c) and 2(c). As the physical pressure on MoTe$_{2}$ increases, $T_{\rm s}$ decreases monotonically, unlike the case for the chemical pressure, under which $T_{\rm s}$ remains almost the same. In addition, as compared with the chemical pressure, the physical pressure enhances $T_{\rm c}$ by a larger extent at the vicinity of $P_{\rm c}$.

\begin{figure}[t]
\begin{center}
\includegraphics[width=7cm]{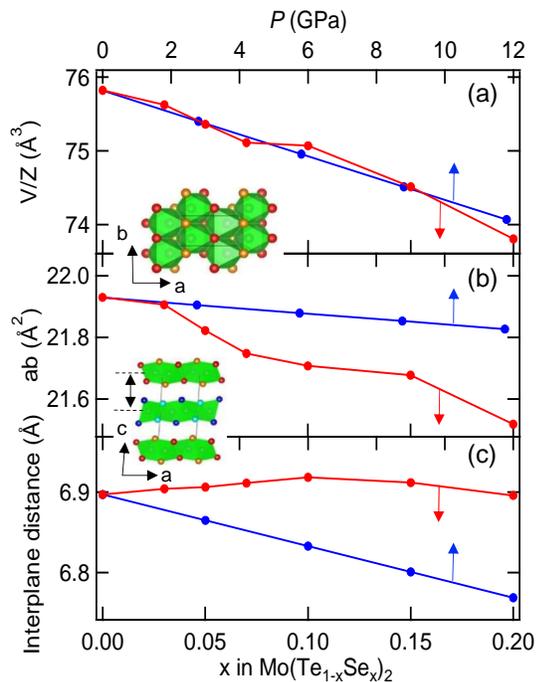}
\caption{(color online). (a) Unit cell volume $V$ divided by number of atoms per unit cell $Z$, (b) $ab$ plane area ($ab$), and (c) interplane distance as a function of $x$ for 1T'-Mo(Te$_{1-x}$Se$_{x}$)$_{2}$ and physical pressure $P$ for 1T'-MoTe$_{2}$. The lattice parameters as a function of $P$ are taken from Ref. \:\onlinecite{16}.}
\end{center}
\end{figure}

To elucidate the possible origin of the difference between the two phase diagrams, we compare the lattice parameters of MoTe$_{2}$ as functions of physical and chemical pressures. As shown by red circles in Fig. 4(a), the unit cell volume per formula unit ($V/Z$) for Mo(Te$_{1-x}$Se$_{x}$)$_{2}$ decreases linearly with increasing $x$, indicating that the Se substitution for Te acts as the chemical pressure. Here, the theoretical unit cell volume under physical pressure calculated in Ref. \:\onlinecite{16} is superimposed on that under chemical pressure, so that the effective pressure by the Se substitution can be estimated as, for instance, 6 GPa for $x=0.1$ (see the horizontal axes at the top and the bottom of Fig. 4). In the phase diagram of Mo(Te$_{1-x}$Se$_{x}$)$_{2}$, the T$_{\rm d}$ phase is completely suppressed by the Se substitution with $x$ somewhere in between 0.05 and 0.1. Thus $P_{\rm c}$ for the chemical pressure is estimated to be in the range 3-6 GPa, which is significantly larger than $P_{\rm c}$ ($<$ 1 GPa) for the physical pressure. This result indicates that the physical pressure is more effective to the suppression of the polar structural transition, which can be discussed from the viewpoint of the anisotropic lattice change. As shown in Fig. 4(b), the $ab$ plane area decreases steeply with increasing the chemical pressure, unlike the case for physical pressure. On the other hand, the interplane distance is strongly compressed not by the chemical pressure but by the physical pressure [Fig. 4(c)]. The large contraction of the interplane distance reflects the weaker interplane coupling as compared with the in-plane coupling with strong covalency in the layered transition-metal dichalcogenides\cite{31}. Considering the difference in the two phase diagrams and the lattice parameters, it is presumable that not only the modification of the band structure but also the enhancement of the three-dimensionality by the interlayer contraction may contribute to the enhancement of $T_{\rm c}$ around $P_{\rm c}$. It has been pointed out that the interlayer contraction increases the hybridization between the Te $p_{\rm z}$ orbitals, giving rise to the enhancement of the three-dimensionality \cite{39}. Given that the driving force for the polar lattice distortion is the enhancement of the interlayer hybridization which reduces the total electron kinetic energy, the physical pressure should suppress the energy gain associated with the polar distortion, leading to the decrease in $T_{\rm s}$ as seen in Fig. 1(c). Further supporting this assumption, the chemical pressure has little impact on the interlayer distance and $T_{\rm s}$ as seen in Figs. 4(c) and 2(c), respectively.

In conclusion, we establish the structural and superconducting phase diagrams of MoTe$_{2}$ as functions of temperature and the physical or chemical pressures. Both kinds of pressure, especially the physical pressure, significantly enhance $T_{\rm c}$ in the vicinity of the critical pressure $P_{\rm c}$, where the polar lattice instability disappears. From the anisotropic lattice changes under the physical and chemical pressures, the interlayer coupling is found to be a key parameter dominating both the structural and superconducting instability. Note that the chemical pressure, which has less influence on the interlayer coupling, enhances $T_{\rm c}$ as well as the physical pressure, implying that not only the interlayer distance but the polar lattice distortion has an impact on the superconductivity. The heat capacity and transport measurements on Mo(Te$_{1-x}$Se$_{x}$)$_{2}$ imply that the electron-phonon coupling is enhanced upon the polar-to-nonpolar transition. This work demonstrates the strong anticorrelation between the polar lattice instability and superconductivity in the Weyl semimetal candidate MoTe$_{2}$.

\section{ACKNOWLEDGMENTS}
The authors appreciate Y. Tokura and K. Ishizaka for their helpful suggestions.
This work is partly supported by JSPS, KAKENHI, Young Scientists B (No. 16K17736), JST PRESTO (Hyper-nano-space design toward Innovative Functionality), Asahi Glass Foundation and Grant for Basic Science Research Projects of the Sumitomo Foundation.


\end{document}